\documentstyle[12pt]{article}

\def\cmm2{{\,\rm cm^{-2}}}
\def\cm2{{\,{\rm cm}^2}}
\def\cmm3{{\,{\rm cm}^{-3}}}
\def\gcmm3{{\,{\rm g\,cm^{-3}}}}

\def\m{\hat{m}}
\def\n{\hat{n}}
\def\l{\hat{l}}

\def\fun#1#2{\lower3.6pt\vbox{\baselineskip0pt\lineskip.9pt
  \ialign{$\mathsurround=0pt#1\hfil##\hfil$\crcr#2\crcr\sim\crcr}}}

\begin{document}
\thispagestyle{empty}

\begin{center}

\vspace{0.3in}
{\Large\bf The  Angular Bispectrum of \\
 The Cosmic Microwave Background}\\

\vspace{.3in}
{\large\bf Xiaochun Luo  } \\
\vspace{0.2in}

{ Center of Particle Astrophysics\\
301 Le Conte Hall\\
 The University of California, Berkeley, CA  94720}\\

\end{center}

\vspace{.3in}
\centerline{\bf ABSTRACT}
\vspace{0.3in}
 COBE has provided us with a whole-sky map of the CBR anisotropies. However,
 even if the noise level is negligible when the  four year COBE data are
available, the cosmic variance
will prevent us from obtaining
information about the Gaussian nature of the primordial fluctuations.
 This important issue is addressed here
by studying the angular bispectrum of the cosmic microwave background
anisotropies.
A general form of the angular bispectrum is given and  the cosmic
variance of the angular bispectrum for Gaussian fluctuations
is calculated. The advantage of using the angular bispectrum is that
one can choose to use the
multipole moments which minimize the cosmic variance term.
 The non-Gaussian signals in most physically
motivated non-Gaussian models are  small
compared with cosmic variance. Unless the amplitudes are  large,
 the non-Gaussian signals are only detectable in the COBE
data  in  those models where the
angular bispectrum is flat or increases with  increasing
  multipole moment.

{\it Subject headings}: cosmic microwave background --- cosmology: theory
\newpage
The cosmic microwave background radiation (CBR) provides one of the most useful
tools for studying the primordial fluctuations that seed  large scale
structures today. Valuable information about the physical processes
that generate  primordial fluctuations in the early universe
can be obtained by studying the statistics of  temperature anisotropies
in the  CBR: are they  Gaussian or non-Gaussian?  Cosmic inflation (Guth 1981;
Linde 1982; Albrecht \& Steinhardt 1982; Liddle \& Lyth 1993)
 predicts a Gaussian pattern of anisotropies on all angular
scales. Spontaneous symmetry breaking, on the other hand, will lead to the
formation of topological defects, and the   pattern of temperature
anisotropies produced by defect-networks tends to be non-Gaussian
(Bouchet et al. 1988; Turok \& Spergel 1991; Turner et al. 1991).
After COBE's detection of CBR anisotropies on large angular scales (Smoot et
al.  1992), both skewness and the three point temperature
 correlation function (which
contain the lowest order deviations from a Gaussian) were proposed
to test  the Gaussian nature of the CBR (Luo \& Schramm
 1992, 1993; Falk et al. 1993). However, there is a problem in applying these
 statistics to test for Gaussianity in the COBE map even with all four years
of
 data available (Hinshaw et al. 1993): namely the  cosmic variance
(Abbot \& Wise 1984; Scaramella \&
Vittorio 1990, 1991; Cayon et al. 1991;  White et al. 1993), i.e.
the fact that one can only
measure CBR temperature anisotropies in one single universe, which
  introduces
theoretical uncertainty in the  skewness and three point function. Theoretical
studies (Srednicki 1993) and Monte-Carlo simulations (Scaramella \&
Vittorio 1990, 1991) show that the cosmic variance is large.
These results cast doubt on whether one could ever
get any information about the non-Gaussian nature of CBR from large scale
experiments such as COBE,  simply because of the limitations from cosmic
variance. The goal of this paper is to address this important issue in the
context
of the angular bispectrum.  We
  will introduce and discuss several properties of  the angular bispectrum,
including the relationship to the bispectrum of the  perturbations
of the gravitational potential, our  modeling of the angular bispectrum,
  and
the cosmic variance of the angular bispectrum when the fluctuations are
Gaussian. Limits on model parameters and the observability of the angular
bispectrum are also discussed.

\section{Angular Power Spectrum}
Before we move on to discuss the angular bispectrum, let us briefly review
 two point statistics first. In the case when the temperature anisotropies are
Gaussian, the
two-point statistic is  all we need to quantify the temperature anisotropy
 patterns
observed in the experiments (such as COBE).

The CBR temperature anisotropy is a 2D random field defined on the two-sphere.
One can perform a spherical harmonic expansion of the temperature
anisotropy on the sky:
\begin{equation}
{\Delta T \over T_{0}} (\theta, \phi) = \sum_{lm} a_{l}^{m} Y_{l}^{m} (\theta,
\phi).
\end{equation}
Then the spherical harmonic expansion coefficients $a_{l}^{m}$, $l\neq 0$,
are random variables, and their statistics (usually called angular statistics)
will completely specify the statistics of the temperature anisotropy itself.
The angular power spectrum $C_{l}$,
 is the $lth$ component of the Legendre
polynomial expansion of the two-point temperature correlation
function (Bond and Efstathiou, 1987):
\begin{equation}
C(\theta) = \left\langle{\Delta T (\vec{m})\over T_{0}} \cdot {\Delta T
(\vec{n})\over T_{0}}\right\rangle = {1\over 4\pi} \sum_{l} (2 l +1) C_{l}
P_{l} (\cos\theta).
\end{equation}
Here the angled brackets denote an ensemble average and
$C_{l}$ is related to the spherical harmonic expansion coefficients through:
\begin{equation}
\langle{a_{l}^{m} a_{l^{'}}^{m^{'}*}}\rangle = C_{l} \delta_{ll^{'}} \delta_{m,
m^{'}}.
\label{C}
\end{equation}

In  realistic experimental settings there are two major factors which
 modify the
 theoretical angular power spectrum defined
 in Eq. (\ref{C}). One is the finite
beam effect, which filters all the high
multiple moments. For large scale experiments such as COBE, which have a finite
beam and full sky coverage, the angular power spectrum is modified as (Silk \&
Wilson 1980; Bond \& Efstathiou 1984):
\begin{equation}
\tilde{C}_{l} = C_{l} W_{l} = C_{l} \exp ( - l (l +1) \sigma^{2}),
\end{equation}
where $\sigma $ is the Gaussian
beam width.  The other important factor is that one can only
measure CBR fluctuations  in a single universe (our own universe),
 thus the observed $C_{l}^{\rm obs}$  is still
a stochastic quantity, which has  intrinsic fluctuations. Here
 $C_{l}^{\rm obs}$ is  the rotationally invariant sum
\begin{equation}
C_{l}^{\rm obs} = {1\over 2 l +1} \sum_{m=-l}^{l} a_{lm} a_{lm}^{*}.
\end{equation}
The cosmic variance of $C_{l}^{\rm abs}$ is then given by:
\begin{equation}
\sigma^{2}_{l} = \langle(C_{l}^{\rm obs})^{2}\rangle - \langle(C_{l}^{\rm
obs})\rangle^{2} = {2\over 2 l +1} C_{l}^{2}.
\end{equation}
 For
a Harrison-Zeldovich (H-Z) primordial spectrum, $C_{l}$ is given by (Peebles
1982)
\begin{equation}
{C_{l}\over 4\pi} = {{6} Q^{2} \over 5 l (l +1)},
\end{equation}
where $Q$ is the rms ensemble-averaged quadruple. For large $l$,
$\sigma_{l}^{2} \propto l^{-5}$ is a rapid decreasing
function of $l$. The cosmic variance  dominates the lower multipole
moments,
 and for high $l$-multipoles it is negligible. We expect this is also true
for the cosmic variance of the skewness and three point function, and we would
expect that the theoretical uncertainties due to cosmic variance will be
reduced  by subtracting out lower multipole moments. We will return to
this point in section 3.

\section{Angular Bispectrum}

If the CBR is Gaussian,  the odd moments of the spherical harmonic
coefficients $a_{l}^{m}$ will vanish. Thus, the non-vanishing odd moments
will be a clear signature of deviation from Gaussianity. The lowest
order of such moments,
 the angular bispectrum, $B_{3}(l_{1}m_{1},
 l_{2}m_{2}, l_{3}m_{3})$, is defined as:
\begin{equation}
B_{3}(l_{1}m_{1}, l_{2}m_{2}, l_{3}m_{3}) \ \ = \ \ \langle a_{l_{1}}^{m_{1}}
a_{l_{2}}^{m_{2}}
a_{l_{3}}^{m_{3}} \rangle.
\end{equation}
The three point temperature correlation is related to the
angular bispectrum through:
\begin{eqnarray}
\xi (\l, \m,\n) = \langle {\Delta T\over T_{0}} ({\l}) {\Delta T\over T_{0}}
({\m})
 {\Delta T\over T_{0}} ({\n})\rangle = \nonumber\\
  \sum_{l_{i}, m_{i}} B_{3}(l_{1}m_{1}, l_{2}m_{2}, l_{3}m_{3})
 Y_{l_{1}}^{m_{1}} (\l) Y_{l_{2}}^{m_{2}}(\m) Y_{l_{3}}^{m_{3}}(\n).
\end{eqnarray}
Because  the three point function is rotationally invariant, the
angular bispectrum $B_{3}(l_{1}m_{1}, l_{2}m_{2}, l_{3}m_{3})$ is non-zero
only if $l_{i}m_{i}, i=1, 2, 3$ satisfy the following conditions \footnote{
These conditions can be derived by choosing special beam configurations.
Condition (1) is a result of the rotational invariance of $\xi$ when two
beams coincide. Condition (2) results from the invariance of $\xi$
under spatial inversion: $\l \rightarrow -\l, \ \ \m  \rightarrow
-\m, \ \ \n \rightarrow -\n$. Condition (3) results from choosing beam
configurations so that the rotational axis is perpendicular to the plane
defined by the three beam directions.}:

(1) $l_{1}, l_{2}, l_{3}$ satisfy the triangle rule, i.e. $l_{i} \leq |l_{j} -
l_{k}|$,

(2) $l_{1} + l_{2} + l_{3} = $ even, and

(3)  $m_{1} + m_{2} + m_{3} = 0$.

The angular bispectrum can be calculated for any given non-Gaussian model.
In this paper we consider only the case where the perturbations are adiabatic,
 so that the temperature anisotropy is related to the gravitational $\phi$ at
the last scattering surface through the
Sachs-Wolfe formula (Sachs \& Wolfe 1967),
\begin{equation}
{\delta T \over T} = {\phi \over 3}.
\end{equation}
The three point temperature correlation function
is related to the bispectrum of the gravitational potential $\phi$,
$P_{\phi} (k_{1}, k_{2}) \equiv \langle \phi_{k_{1}} \phi_{k_{2}}
\phi_{k_{3}} \rangle
({\bf k_{1} + k_{2} + k_{3}} = 0)$, through
\begin{equation}
\xi_{T} (\l, \hat{m}, \hat{n}) = {1\over 27}\cdot\int
P_{\phi}(k_{1},k_{2}){e^{i\eta_{0}(\hat{k_{1}}\hat{m}+\hat{k_{2}}\hat{n}
+\hat{k_{3}}\hat{l})}}\delta^{3}(\vec{k_{1}}+\vec{k_{2}}+\vec{k_{3}})
{d^{3}k_{1}d^{3}k_{2}d^{3}k_{3}\over(2\pi)^{9}},
\end{equation}
where $\eta_{0} = 2 c/H_{0}$ is the distance to the last scattering surface
and $\hat{l}, \hat{m}, \hat{n}$ are the beam directions.
The general relation between the angular bispectrum $B_{3}$ and the bispectrum
$P_{\phi}$ is rather complicated and we will not give it here. The expression
simplifies if $P_{\phi} (\vec{k}_{1}, \vec{k}_{2})$ depends only upon the
amplitudes of $\vec{k}_{1}$ and $\vec{k}_{2}$. This is what we expect for the
angular bispectrum from non-standard inflationary scenario after proper
symmetrization.
 In this case, by using the following expansions,
\begin{equation}
e^{i\hat{k} \cdot \hat{m} \eta_{0}} = { 4 \pi} \sum_{l} i^{l}  j_{l}
(k\eta_{0})
\sum_{m} Y_{lm} (\hat{k}) Y^{*}_{lm} (\hat{m}),
\end{equation}
and
\begin{equation}
Y_{l_{1}}^{m_{1}} Y_{l_{2}}^{m_{2}}
= \sum_{l_{3}m_{3}} A(l_{1}l_{2}l_{3}, m_{1}m_{2}m_{3}) Y_{l_{3}}^{m_{3}},
\end{equation}
we find that
\begin{equation}
B_{3}
= {4\over 27 \pi^{2} }\int dk_{1} dk_{2} k_{1}^{2}k_{2}^{2} P_{\phi} (k_{1},
k_{2}) j_{l_{1}}^{2} (k_{1})
j_{l_{2}}^{2} (k_{2})  A(l_{1}l_{2}l_{3}, m_{1}m_{2}-m_{3}),
\label{special}
\end{equation}
where
\begin{equation}
A(l_{1}l_{2}l_{3}, m_{1}m_{2}-m_{3}) = \left[{(2 l_{1} +1) ( 2 l_{2} + 1) \over
4\pi ( 2 l_{3}  + 1)}\right]^{1/2}
C(l_{1}l_{2}l_{3}, 000) C(l_{1}l_{2}l_{3}, m_{1}m_{2}-m_{3})
\end{equation}
and $C(l_{1}l_{2}l_{3}, m_{1}m_{2}m_{3})$ are Clebsh-Gordan coefficients.
The expression \newline $A(l_{1}l_{2}l_{3}, m_{1}m_{2}-m_{3})$
 is nonzero only if $m_{1} + m_{2} + m_{3} = 0$, $l_{1}
+ l_{2} + l_{3}$ = even and
$l_{1}, l_{2}, l_{3}$ satisfy the triangle rule.
 Thus, the rotational
invariance of the three point function is guaranteed.

Let us recall that the angular power spectrum $C_{l}$ is related to the power
spectrum
$P_{\phi}(k)$  of fluctuations in $\phi$ through (Kolb \& Turner 1990)
\begin{equation}
C_{l} = {2\over 9\pi} \int dk k^{2} P_{\phi}(k) j_{l}^{2}(k\eta_{0}).
\label{cl}
\end{equation}
The functional form of $B_{3}$ given by Eq.(\ref{special}) and Eq.(\ref{cl})
 motived us
to  model the angular bispectrum as the following:
\begin{eqnarray}
B_{3} (l_{1}m_{1}, l_{2}m_{2}, l_{3}m_{3})
=  C^{3/2}(0) A(l_{1}l_{2}l_{3}, m_{1}m_{2}-m_{3}) \times \nonumber \\
\{\alpha [\bar{C} (l_{1}) \bar{C} (l_{2}) + \bar{C} ({l_{2}})
\bar{C}({l_{3}}) +
\bar{C} (l_{3}) \bar{C} ({l_{1}})] + \beta [\bar{C} (l_{1}) \bar{C} (l_{2})
\bar{C} (l_{3})]^{\gamma} \},
\label{bi}
\end{eqnarray}
where $\alpha, \beta, \gamma$ are three dimensionless
parameters,  $\bar{C}({l}) = C_{l} /C(0)$
is the normalized angular power spectrum and $C(0) = {1\over 4\pi}\sum
\tilde{C}_{l} (2 l + 1)$. For COBE,  where the FWHM beam width
is $7^{\circ}$, $C(0) = 4.63 Q^{2}$ and  $\bar{C}({l})
= {3.26\over l (l +1)}$.

Apart from guaranteeing the rotational invariance of the three point
function, the bispectrum modeled by Eq.(\ref{bi}) has another advantage:
several physically motivated non-Gaussian scenarios give distinctive
predictions for the  constants $\alpha, \beta,  \gamma$.
These constants can be calculated directly from  Eq.(\ref{special})
and Eq.(\ref{cl}) once the perturbation
bispectrum  $P_{\phi} (k_{1}, k_{2})$ is given for a specific
non-Gaussian model. In practice, it was found that  in several
non-Gaussian models of cosmological interest, it is easier to derive an
expression for the three point correlation function $\xi(\hat{r}_{1},
\hat{r}_{2},\hat{r}_{3})$ in terms
of the two point function $C(\theta)$ than to find an expression for
$P_{\phi}(k_{1}, k_{2})$ (Luo 1993).
It is convenient to use the normalized two point functions $\psi$ and three
point function $\eta$, where
\begin{equation}
\psi (|\hat{r}_{1} - \hat{r}_{2}|) = C_{2} (\hat{r}_{1}, \hat{r}_{2})/C_{2}(0),
\  \ \psi (0) = 1,
\end{equation}
\begin{equation}
\eta (\hat{r}_{1}, \hat{r}_{2}, \hat{r}_{3}) = \xi(\hat{r}_{1},
\hat{r}_{2},\hat{r}_{3}) /C_{2} (0)^{1.5}, \  \ \eta(0) = \mu_{3},
\end{equation}
and  $\mu_{3}$ is the skewness.
The theoretical predictions for three point function in specific scenarios
are the following:

\noindent
{\bf (1) Inflation}

Various non-standard inflation models
will generate a non-zero three point correlation function
(Falk et al. 1993; Luo 1993; Gangui et al. 1993). The generic form of the three
point function in most inflationary models is
\begin{equation}
\eta (\hat{r}_{1}, \hat{r}_{2}, \hat{r}_{3}) = {\mu_{3}\over 3}
(\psi (|\hat{r}_{1} - \hat{r}_{2}|) \psi (|\hat{r}_{1} - \hat{r}_{3}|) +
\psi (|\hat{r}_{1} - \hat{r}_{2}|) \psi (|\hat{r}_{2} - \hat{r}_{3}|) +
\psi (|\hat{r}_{1} - \hat{r}_{3}|) \psi (|\hat{r}_{2} - \hat{r}_{3}|)),
\end{equation}
where $\mu_{3}$ is a dimensionless constant.
For slow-roll inflation models with one field and a cubic self-coupling,
$\mu_{3} \sim 10^{-6}$ ( Falk et al. 1993).  Non-linear gravitational
evolution  will produce a three point function of similar form with
$\mu_{3} \sim 0.01$ (Luo \& Schramm 1993).

\noindent
{\bf (2) $\chi^{2}_{n}$ fields}

Consider the $\chi^{2}_{n}$ field $ Y = \sum_{i=1}^{n} X_{i}^{2}$, which
 describes the $O(N) \sigma$ model of global topological
defects (Turok \& Spergel 1991) in the large $ N$ limit, with $n = 4N$.
The three point function is found to be
\begin{equation}
\eta (\hat{r}_{1}, \hat{r}_{2}, \hat{r}_{3}) = \sqrt{2/N} (\psi (|\hat{r}_{1} -
\hat{r}_{2}|) \psi (|\hat{r}_{1} - \hat{r}_{3}|) \psi (|\hat{r}_{2} -
\hat{r}_{3}|))^{3/2}.
\end{equation}

\noindent
{\bf (3) Late-time Phase Transitions}

In this model (Hill et al. 1989), due to the conformal invariance of the system
at the critical point (Polyakov 1970), the three
point function  has the following  form (Luo \& Schramm 1993):
\begin{equation}
\eta (\hat{r}_{1}, \hat{r}_{2}, \hat{r}_{3}) = \alpha (\psi (|\hat{r}_{1} -
\hat{r}_{2}|) \psi (|\hat{r}_{1} - \hat{r}_{3}|) \psi (|\hat{r}_{2} -
\hat{r}_{3}|))^{3},
\end{equation}
where $\alpha$ is a dimensionless constant of order unity.

Once the three point function is found, the angular bispectrum $B_{3}$
and  constants $\alpha, \beta, \gamma$
 can be found from Eq. (9) by expanding
the three point function and two point function in spherical harmonics.
Non-standard
inflation and non-linear gravitational evolution predict $\beta = 0, \alpha =
\mu_{3}/3$, where $\mu_{3}$
is the skewness; Late-time
phase transitions (LTPT)  will predict $\alpha \sim0,
\gamma \sim1$ and $\beta$ of order unity; $O(N)$
 $\sigma$ models  predict
 $\alpha \sim 0, \gamma \sim 1/2$
and $\beta \sim \sqrt{2/N}$ (Luo 1993). Another possibility is that the
underlying CBR fluctuations are non-Gaussian only on
 smaller angular scales. In this case,
the angular bispectrum will increase slowly
 with the increasing of $l$, and  will peak  around $l_{c}$
  corresponding to
 the characteristic scale $\theta_{c}$ where the CBR is highly non-Gaussian.
If $\theta_{c} \sim 1^{\circ}$, then $l_{c} \sim 100$. Since the
observed anisotropies in large angle experiments (such as COBE) are the
beam-smoothed fluctuations, we would expect the angular
bispectrum to be flat, i.e. $\alpha=0, \gamma =0$, at low $l$. Thus, if future
analysis of COBE
map reveals
a non-vanishing angular bispectrum, fitting the experimental results to the
functional form given by Eq.(\ref{bi}) will help to discriminate between
different
non-Gaussian models.

\section {Cosmic Variance of the Angular Bispectrum}
The CBR anisotropies are stochastic in nature, which will give
rise to a theoretical uncertainty in the particular realization of our sky.
This effect is severe in  large
scale CBR experiments, and  limits our ability to extract
information about non-Gaussian nature of CBR from current experimental data.
 In this section we will address
this important issue and calculate the cosmic variance of the bispectrum
when the fluctuations are Gaussian. We will show how to minimize the cosmic
variance  by choosing the appropriate pairs of multipole moments.
In order for the non-Gaussian signals to rise above the cosmic
variance, bounds on the model parameters $\alpha, \beta, \gamma$
of the bispectrum defined in the previous section are also discussed.

The cosmic variance of the angular bispectrum is given by
\begin{equation}
\sigma_{3}^{2} =
\langle a_{l_{1}}^{m_{1}} a_{l_{2}}^{m_{2}} a_{l_{3}}^{m_{3}}
a_{l_{1}}^{m_{1}*} a_{l_{2}}^{m_{2}*} a_{l_{3}}^{m_{3}*}\rangle
\end{equation}
When $a_{l}^{m}$ is Gaussian, by using Wick's theorem,
the expression above reduces to
\begin{eqnarray}
\sigma_{3}^{2} =
C_{l_{1}} C_{l_{2}} C_{l_{3}} +
 8 \delta_{l_{1} l_{2} l_{3}} \delta_{m_{1}m_{2}m_{3},0} C_{l_{1}}^{3}+
 \delta_{l_{1}l_{2}} [\delta_{m_{1},-m_{2}} + \delta_{m_{1}m_{2}}]
C_{l_{1}}^{2} C_{l_{3}} +
\nonumber\\
\delta_{l_{2}l_{3}} [\delta_{m_{2},-m_{3}} + \delta_{m_{2}m_{3}}]
C_{l_{2}}^{2} C_{l_{1}} +
\delta_{l_{3}l_{1}} [\delta_{m_{1},-m_{3}} + \delta_{m_{1}m_{3}}]
C_{l_{3}}^{2} C_{l_{2}}
\end{eqnarray}
where $C_{l} = \langle a_{l}^{m} a_{l}^{m*}\rangle$
 is the angular power spectrum.
{}From the  above, one also finds that the variance of $B_{3}$ depends
heavily upon the choice of multipole moment pairs $(l_{i}m_{i})$.
The variance of the terms where $l_{1} \neq l_{2} \neq l_{3}$ is given by
\begin{equation}
\sigma_{3}^{2} = C_{l_{1}} C_{l_{2}} C_{l_{3}},
\end{equation}
while  for $l_{1} =l_{2}=l_{3}={l}, \ \ m_{1}=m_{2}=m_{3}=0$, it is given by
\begin{equation}
\sigma_{3}^{2} = 15 C_{l_{1}} C_{l_{2}} C_{l_{3}} = 15 C_{l}^{3}.
\end{equation}
The variance of the bispectrum can differ by more than one order of magnitude
by choosing different multipole pairs. It is easy to see why the cosmic
variance of skewness is large:
by writing it in terms of the angular bispectrum, the variance of
the skewness is given by
\begin{equation}
\eta^{2} \approx \sum_{l_{i}m_{i}} \sigma_{3}^{2}(l_{1}m_{1}, l_{2}m_{2},
l_{3}m_{3}) A^{2}(l_{1}l_{2}l_{3}, m_{1}m_{2}m_{3}).
\end{equation}
This is a sum over all different multipole moments, and  is dominated
by  terms such as $m_{i} =0, l_{1}=l_{2}=l_{3}=l$. The advantage of
using the angular bispectrum is that one can choose multipole moment
pairs to minimize the cosmic variance, and at the same time, to
maximize model predictions for the angular bispectrum.

One way to reduce cosmic variance is to subtract out the lower multipole
moments. Part of the CBR
signals will then
also be removed so that the signal-noise ratio will be reduced.
However, this will not be a serious problem for COBE map with fours years
of data. The noise levels in COBE maps diminish
rapidly ($\propto {\rm time}^{-3/2}$) with additional data (Hinshaw et al.
1993).
In the four year maps, the noise level will be roughly eight times smaller
than in the first year skymap. For a scale invariant H-Z initial fluctuation
spectrum, the CBR signal is reduced by merely a factor of $3$ by removing the
quadrupole. Thus, the signal-noise ratio in the quadrupole-removed four-year
map will still be a factor of
three better than the first year map (with quadrupole).

 A more serious problem with subtracting out lower moments is that part of the
non-Gaussian signal will also be removed. The amount of
non-Gaussian signal
removed is model-dependent.    We list in Table 1 the
angular bispectrum $B_{3}$, modeled by Eq. (\ref{bi}),
 in various non-Gaussian models,
compared with the cosmic variance $\sigma_{3}$.  In non-standard inflation
or late-time phase transition models, most of the non-Gaussian signal
will be removed by subtracting out the lower multipole
moments. In these types of model, one should focus on the lowest multipole
moment ($l=2$). The low limits on the model constants are: $\alpha > 0.6$
in non-standard inflation, where $\beta = \gamma=0$; $\beta > 3.2$ for
late-time
phase transitions, where $\alpha=0, \gamma=1$. In models where $\alpha =0,
\gamma=1/2$, the ratio of non-Gaussian signal to the cosmic variance
is approximately the same for the low $l$ moments, but since the instrumental
noise is relatively high for large moments, one will also lose information
by subtracting out lower moments. The low limit on $\beta$ is $\beta > 1.5$.
 For most
physically motivated models (non-standard inflation,
non-linear gravitational evolution, the late-time phase transition
or the $O(N) \ \ \sigma-$models) where the model constants $\alpha, \beta$
are less than unity, all terms of the angular bispectrum are  smaller
than the cosmic variance term.
 Since the cosmic variance term
 $\sigma_{3} \propto l^{-3}$  for  the H-Z initial fluctuation spectrum,
if the angular bispectrum falls
off faster than $l^{-3}$, it will be impossible for the experiments
to detect  deviations from Gaussian behavior, unless the amplitude of the
non-Gaussian signature is unrealistically large. This suggests
that the decisive test of the Gaussian nature of fluctuations has to come from
degree scale experiments, as we have stressed before (Luo 1993).
Nevertheless, in models where the angular bispectrum is flat ($\gamma =0$),
even a very small deviation from Gaussianity ($\beta \sim 0.02$)
can stand out above the cosmic variance, by analyzing the angular bispectrum
at higher multipole moments ($l \sim 8$).

\bigskip
We wish to thank D.N. Schramm,  J. Silk, M. White, D. Scott,
M. Srednicki and anonymous refree
 for helpful comments and suggestions. This
work was supported in part by DoE and NSF.
\newpage

\def\ref{\par\noindent\hangindent=2pc \hangafter=1 }
\centerline{REFERENCES}

\bigskip

\ref
Abbott, L.F., \& Wise, M.B. 1984, ApJ, 282, L47

\ref
Albrecht, A., \& Steinhardt, P. 1982, Phys. Rev. Lett. 48, 1220

\ref
Bond J. R., \& Efstathiou, G. 1984, ApJ, 285, L45

\ref
Bond J. R., \& Efstathiou, G. 1987, MNRAS, 226, 655

\ref
Bouchet, F.R., Bennett, D., \& Stebbins, A.J. 1988, Nature, 335, 410

\ref
Cayon, L., Martinez-Gonzalez, E., \& Sanz, J.L., 1991, MNRAS, 253, 599

\ref
Falk, T., Rangarajan, R., \& Srednicki, M. 1993, ApJ, 403, L1

\ref
Gangui, A., Lucchin, F., Matarrese, S., \& Mollerach, S. 1993,
SISSA REF 193/93/A, astro-ph/9312033

\ref
Guth, A.H. 1981, Phys. Rev. D23, 347

\ref
Hill, C.T., Schramm, D.N., \& Fry, J. 1989, Comm. Nucl. Part. Phys., 19, 25

\ref
Hinshaw G. et al. 1993, COBE Preprint 93-12

\ref
Kolb, E.W., \& Turner, M.S. 1990, The Early Universe (Redwood City: Addison
Wesley)

\ref
Liddle, A., \& Lyth, D. 1993, Phys. Rep., 231, 1

\ref
Linde, A. 1982, Phys. Lett. B 108, 389

\ref
Luo, X.C. 1993, Fermilab-Pub-93/294-A, Phys. Rev. D in press

\ref
Luo, X.C., \& Schramm, D.N. 1993, ApJ, 408, 33

\ref
Luo, X.C., Schramm, D.N. 1993, Phys. Rev. Lett. 71, 1124

\ref
Peebles, P.J.E. 1982, ApJ, 263, L1

\ref
Polyakov, A.M. 1970, JEPT Lett., 12, 538

\ref
Sachs, R.K., \& Wolfe, A.M. 1967, ApJ, 147, 73

\ref
Scaramella, R., \& Vittorio, N. 1990, ApJ, 353, 372

\ref
Scaramella, R., \& Vittorio, N. 1991, ApJ, 375, 439

\ref
Silk, J., \& Wilson, M.L. 1980, Physica Scripta, 21, 708

\ref
Smoot, G. et al. 1992, ApJ, 396, L1

\ref
Srednicki, M. 1993, ApJ, 416, L1

\ref
Turok, N., \& Spergel, D.N. 1991, Phys. Rev. Lett., 66, 3093

\ref
White, M., Krauss, L. \& Silk, J. 1993, ApJ, 418, 535

\bigskip
\centerline{TABLE CAPTION}

\bigskip
\noindent
Table 1: Comparison of the angular bispectrum $B_{3}$ in various
non-Gaussian models with the cosmic variance of the angular bispectrum
 for
 Gaussian fluctuations, weighted by $A(l_{i}, m_{i})$. The multipole pairs
($l_{i},m_{i}$) are chosen
to obtain the maximum value for $B_{3}$ in non-Gaussian models,
and at the same time, to keep the cosmic variance as small as possible.

\newpage
\vskip 0.1 in
\centerline{Table 1}
\vskip 0.2 in
\begin{tabular} {|c|c|c|c|c|c|}
\hline
\hline
 & Gaussian:  &
\multicolumn{4}{c|} {Non-Gaussian Models:} \\
&  $B_{3} =0$& \multicolumn{4}{c|} {$l^{3} B_{3}(l_{i}m_{i})/C^{3}(0)$} \\
\cline{2-6}
Multipole & Cosmic & Inflation & LTPT & $O(N) \sigma$ & Smoothing \\
pairs:&variance: & $\alpha\neq 0$ & $\alpha=0$& $\alpha=0$ & $\alpha =0$ \\
$(l_{1}l_{2}l_{3},m_{1}m_{2}m_{3})$ &${l^{3} \sigma_{l} A(l_{i}m_{i})\over
C^{3}(0)}$  & $\beta=0$ &$\gamma=1$  &
$ \gamma=1/2$& $  \gamma=0$ \\ \hline
(222, 11-2) & 1.00 &1.56 $\alpha$ & 0.28 $\beta$ & 0.70 $\beta$ & 1.77 $\beta$
\\ \hline
(444, 22-4) & 1.14 & 0.97 $\alpha$ & 0.05 $\beta$ & 0.80 $\beta$ & 12.2 $\beta$
\\ \hline
(666, 33-6)&1.15 & 0.70 $\alpha$ & 0.02 $\beta$ & 0.81 $\beta$ & 37.6 $\beta$
\\ \hline
(888, 44-8) & 1.13 & 0.51 $\alpha$ & 0.007 $\beta$ & 0.80 $\beta$ & 83.4
$\beta$ \\ \hline
(101010,55-10) &1.11 & 0.40 $\alpha$ & 0.004 $\beta$ & 0.76 $\beta$ & 154.1
$\beta$ \\ \hline
\hline
\end{tabular}

\end{document}